\documentclass{rspublic}
\usepackage[dvips]{epsfig}

\begin{document}

\title[Magnetic Fields In Astrophysical Objects ]{Magnetic Fields In Astrophysical Objects}

\author[L. J. Silvers]{L. J. Silvers}

\affiliation{Department of Applied Mathematics and Theoretical
Physics, University of Cambridge, Wilberforce Road, Cambridge,
Cambridgeshire, CB3 0WA, U.K.}

\label{firstpage}

\maketitle

\begin{abstract}{Magnetic Fields -- Magnetohydrodynamics (MHD) --
    Accretion Discs -- Stellar and Planetary Dynamos -- Planetary
    Magnetic Fields}

Magnetic fields are known to reside in many astrophysical objects and are now believed to be
 crucially important for the creation of phenomena on a wide variety of scales. However, the role of the magnetic field in the bodies that we observe has not always been clear. In certain situations, the
importance of a magnetic field has been over looked on the grounds that the large-scale
magnetic field was believed to be too weak to play and important role in the dynamics. 

In this article I discuss some of the recent developments concerning
magnetic fields in stars, planets and accretion discs. I choose to
emphasise some of the situations where it has been suggested that weak
magnetic fields may play a more significant role than previously thought.  At the end of the article I list some of the questions to be answered in the future.

\end{abstract}

\section{Introduction}
Our knowledge of magnetism, and of magnetic fields, began with the
study of lodestones several hundred years before Christ. Stones of
this type can become magnetic, and then two such stones will
naturally align (see, for example, Parker 1979). Scholars of the time proposed a
variety of reasons for this peculiar behaviour and it was even
suggested at one point that the lodestone might even have a soul!

Many centuries later, at the start of the $17^{th}$ century, William
Gilbert noted that the Earth also behaves like a large lodestone
(see for example Childress \& Gilbert, 1995). Gilbert noted that a
lodestone always points to align with what is now referred to as
magnetic north/south in the same way that a small stone aligns
itself relative to a larger one. By the end of the nineteenth
century, it had been concluded that the Earth is not unique in having a
magnetic field. In 1908, Hale  determined that the Sun has
a magnetic field and that the phenomena that are known as sunspots, which
are dark patches on the surface of the Sun (Hale 1908), have a magnetic field
that is incredibly strong (of the order of 1000 Gauss). In the last
century, studies of different stars, planets and other
non-Earth objects, have been made and it is now known that many of them have a
detectable magnetic field. A useful summary of typical field strengths for some objects can be
found in Zeldovich \textit{et al}.\ (1983) and Jones (2007).

Given that there is a measurable magnetic field in many objects
in the Universe, it is natural for us to ask what its role is in the
dynamics of different objects.  At a more fundamental level we wish
to know if the
magnetic field that is being measuring today is a primordial field,
that has resided in the object since its creation, or if the
magnetic field is one that is constantly being generated by some kind of
hydromagnetic dynamo mechanism.

In the last few decades, the importance of a magnetic field in many
of the astrophysical bodies has become increasingly
recognised. This has stimulated considerable research in the area of
magnetohydrodynamics (MHD) to understand how a plasma and a magnetic
field interact. In some phenomena, such as sunspots, the role of
such a  strong magnetic field is fairly easy to grasp at a basic
level but only recently has a more comprehensive
understanding of these phenomena been obtained (see the article by Bushby in this
volume). However, it has proved to be more difficult to understand the important and role of weaker
fields, where the magnetic energy is much less than the kinetic energy
\footnote{This is the definition of weak that is frequently used for
  astrophysical scenarios. However, we note here that is some cases
  the definition of a weak magnetic field is different and is one for
  which the gas pressure is much bigger than the magnetic pressure.}. Further, understanding why some
planets have magnetic fields that are easy to detect, and others don't means that there is a
vast array of issues and interesting questions still to be
resolved
about magnetic fields and their interactions with electrically
conducting fluids in objects throughout the universe.

In this short article, I will discuss some of the recent
developments in our knowledge and understanding of astrophysical
magnetic fields, and discuss some of the interesting open problems
and issues. Much of this revolves around the study of how
astrophysical object generate magnetic fields i.e.\ their dynamo
mechanisms.


\section{The Role of the magnetic field in the Sun, other stars and planets}

\begin{figure}[htbp]
\begin{center}
\epsfig{file=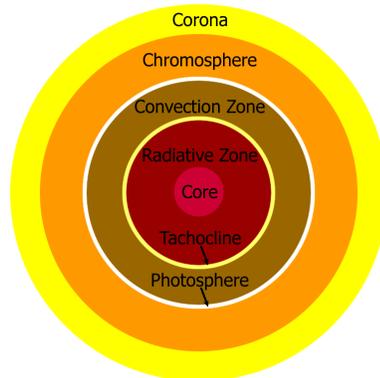,width=5.0cm,height=5.0cm}
\end{center}
\caption{An illustration showing the different zones of the Sun.}
\label{Sun}
\end{figure}

\begin{figure}[htbp]
\begin{center}
\psfig{file=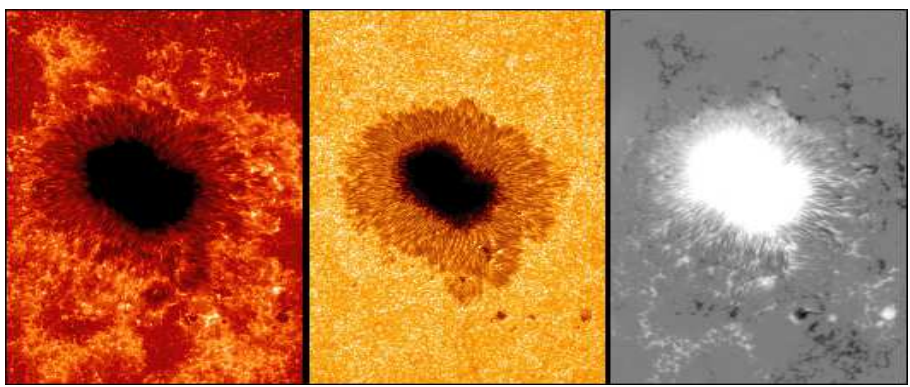,width=12cm,height=5.0cm}
\end{center}
\caption{High resolution (pseudo-colour) images of a sunspot viewed
in
  three different ways. Left: In Ca II H; Middle: in G-band; Right: Magnetogram. Images are courtesy of the Hinode
  mission. (Hinode is a Japanese mission developed and
  launched by ISAS/JAXA, with NAOJ, NASA and
  STFC as partners. It is operated by these
  agencies in co-operation with ESA and NSC.) } \label{Sunspotimage}
\end{figure}

Given that our knowledge of magnetism in non-Earth objects is
greatest for the Sun, this seems a good starting point for our brief
trip around a portion of the magnetic Universe. This star is a
fascinating and complex astrophysical object that has now enthralled
mankind for thousands of years. It is a vast ball of plasma that
can, somewhat crudely, be pictured as being comprised of a series of
layers, as shown in figure \ref{Sun}. Events in the outer parts of
this star (on the the visible surface of the Sun that is referred to
as the photosphere; in the chromosphere and in the corona) can be
observed directly and images such as those from the SOHO mission
show a variety of transient events. Historically, the first solar
phenomena feature to be identified were sunspots, which appear in
the records dating as far back as 350BC and have been known to
persist for several weeks (Tobias 2002). Figure \ref{Sunspotimage}
shows images of a sunspot viewed in three different ways.

In 1908 Hale, utilizing the Zeeman effect (the splitting of a
spectral line in the presence of a magnetic field), was able to show
that these patches on the solar surface are regions with strong
magnetic field. In fact, it is the presence of a strong magnetic
field that inhibits convection and gives rise to spots. They are
dark in appearance as the temperature in a spot is lower than its
surroundings. Recently progress has been made in our understanding of these
phenomena (Thomas, Weiss, Tobias \& Brummell 2002;
2004; 2006) and a non-technical account of this development is
discussed by P.\ J.\ Bushby in this volume (Bushby 2008).

Observations of sunspots lead to the discovery that the Sun's magnetic field has a
cyclic behaviour. At the start of a cycle, sunspots appear in pairs
with the two spots of opposite polarity. These spot pairs are
approximately aligned in a east-west orientation, 30 degrees either
side of the equator  (see, for example, Tobias \& Weiss 2007). The
west-most spot of each pair in the northern hemisphere is always of
the same polarity and opposite to the western spots in the southern
hemisphere (see, for example, Tobias \& Weiss 2007). Over
(approximately) the next 11 years, the latitude at which emergence
of new sunspot pairs gradually moves so that the new pairs in later
years appear closer to the equator. At the end of the 11 years, the
sunspot pairs reappear at higher latitudes but with the polarity of the
spot in the pair reversed. Consequently, the solar cycle does not
repeat every 11 years, but instead on an approximate 22 year period.
It is interesting to note that there have been periods where there
have been no visible spots of the surface of the Sun. The most noted
example in history occurred in the 17th century, and is called the
Maunder minimum (see, for example, Tobias 2002).

While sunspots were the first solar phenomena to be observed, there
are a host of others, which include flares (regions of high
brightness associated with a complex topology of the magnetic field)
and prominences (regions where cool, dense material is suspended
above the surface of the Sun). The origin of the solar magnetic
field, which is crucial for observed transient phenomena, is rooted
deep inside the Sun and therefore the interactions in the solar
interior (below the photosphere), between the plasma and the magnetic field, give rise to
what is observed on, and above, the surface.  Since the observations
of the sunspots and other phenomena change with time, the magnetic
field in the interior must also be time dependent. In order to
produce such a complicated collection of magnetic phenomena, the
interaction between the magnetic field and the plasma must be
complex, and one of the most crucial questions in solar physics is:
\textit{How do a plasma and a magnetic field interact in the
interior of the Sun? }

The solar interior can be divided into three
principal, large, regions -- the core, radiative zone, and
convection zone. Despite the fact that these interior regions are
not observable directly, there is still a fairly large amount that is known about
the internal structure particularly through utilizing a relatively
new branch of solar research called helioseismology. Research in this area
seeks to infer information
about the interior of the Sun via the oscillations
observed at the surface (Stix 1996; Thompson 2004). Helioseismological
inversions (e.g.\ Schou \textit{et al}.\ 1998) suggest that the
differential rotation profile that is observed at the surface of the
Sun is maintained throughout the bulk of the convection zone
Furthermore, the radiative zone rotates essentially as a solid body
with angular velocity equal to that of the surface at a latitude of
approximately $35^\circ$. Thus, separating the convection
and radiative zones, there is a thin transition region of
strong radial shear - a region known as the tachocline. This region
has been postulated for a number of years, but has only recently
been confirmed by helioseismology and, despite its relatively narrow
radial extent, is believed to play a crucially important role in the
evolution of the solar magnetic field due to its large shear.

The solar magnetic field is not believed to be simply a relic field
that has been part of the star since its formation. It is
being generated and maintained by a hydromagnetic dynamo mechanism,
which is thought to be the case partially because of sunspot
observations: the  polarity of the west-most
sunspots changes every 11 years, which suggests the destruction of that
component of the large-scale magnetic field and the generation of that component
of the large-scale magnetic field with the opposite polarity for the next 11
years. As such there has been considerable work to explain how the magnetic field
is being sustained within the Sun. The original concept for the
mechanism that is maintaining the solar dynamo was that it occurred entirely in
the turbulent solar convection zone, as the initial magnetic field
is stretched and folded by the actions of the turbulent flow on it.
However, it was noted in the early 1990s (Parker 1993) that the
entire dynamo mechanism probably does not lie within this region, because a
magnetic field can act back on the flow and impede the rate of
regeneration of the magnetic field (Cattaneo \& Vainshtein 1991; Gruzinov \& Diamond 1996). The crux of the issue is that,
while this effect has always been known to be a function of the
strength of the magnetic field on the flow, it was noted that extremely
weak large-scale magnetic fields can still have an incredibly significant effect
on the flow and so a catastrophic effect on the rate of field regeneration
\footnote{Extremely weak here refers to fields that are such that
  the magnetic energy is $R_m$ times smaller than the kinetic energy,
  where $R_m$ is a measure of the relative importance of the advective
  to diffusive terms that appear in the equation that governs the
  evolution on the magnetic field. $R_m$ is known as the magnetic
  Reynolds number and in the solar convection zone
  $R_m\gg1$.}. This leads to the conclusion that the magnetic field cannot
reside entirely in the solar convection zone, and new mechanisms
such as the interface dynamo mechanism were proposed (starting from
Parker 1993).

\begin{figure}[htbp]
\begin{center}
\epsfig{file=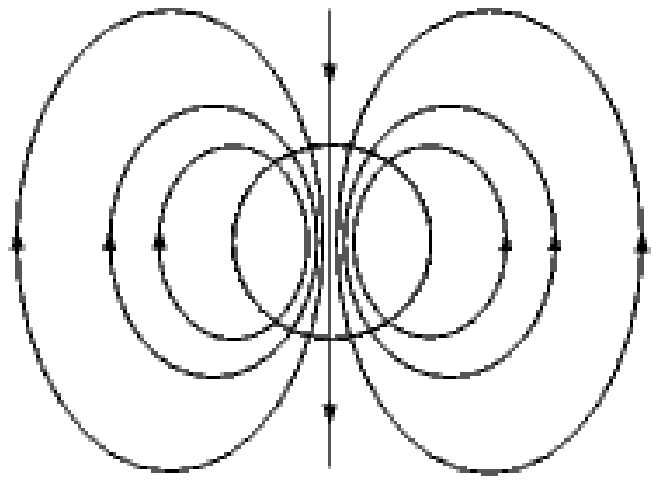,width=5.0cm,height=5.0cm}
\epsfig{file=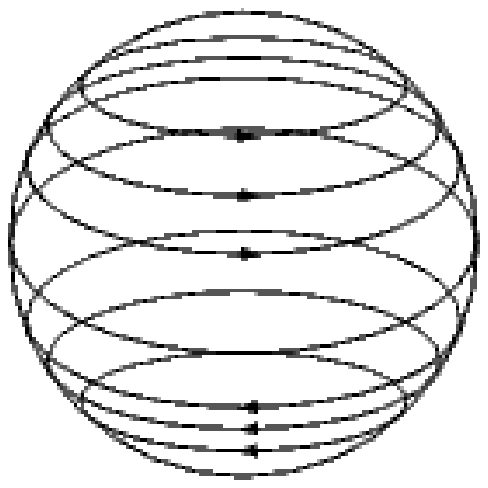,width=5.0cm,height=5.0cm}
\end{center}
\caption{Illustrations showing the poloidal (left) and toroidal (right)
  components of the magnetic field courtesy of J.\ J.\ Love.} \label{PolTor}
\end{figure}

\begin{figure}[htbp]
\begin{center}
\epsfig{file=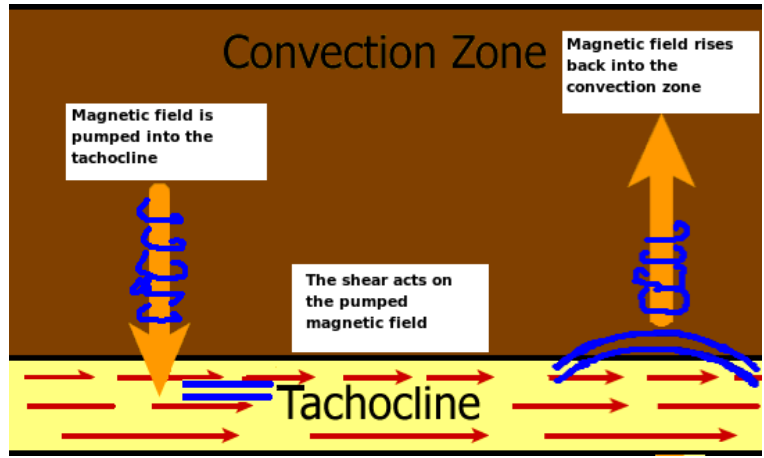,width=10cm,height=6.0cm}
\end{center}
\caption{The interface dynamo mechanism.} \label{interface}
\end{figure}

In the interface model, the turbulence within the convection zone
retains the ability to generate the poloidal component of magnetic
field (see Figure \ref{PolTor}) but it invokes the natural tendency of overshooting convection
to transport the magnetic field out of the convection zone into the
tachocline. This movement of magnetic field out of the convection zone prevents the catastrophic effects mentioned above.
Once in the tachocline the magnetic field is sheared out to generate
a large-scale toroidal component of the field, which then rises returning
magnetic field back into the convection zone for the cycle to
continue. This is shown in figure \ref{interface}.

It is important to note that the interface dynamo mechanism is not the
only current proposal for a mechanism for the solar dynamo and others
include, for
example, the Babcock-Leighton model (Babcock 1961; Leighton
1969)\footnote{In this model the decay of active regions at the solar
  surface releases poloidal field. This component of the magnetic
  field is transported to the base of the solar convection zone by
  meridional circulation where is is sheared out by differential
  rotation to give rise to a toroidal component of the magnetic
  field. The toroidal field then rises to the surface and gives rise
  to 
  sunspots. The decay of these sunspots in active regions then gives
  rise to poloidal field for the cycle to begin again.}. Also, there have been papers in recent years to suggest that
the catastrophic quenching that, in part, motivated the interface
dynamo model may not be as severe as anticipated (Blackman \& Field
2000; Silvers 2006). However, recent observational evidence from another star, $\tau$
Boo, with some characteristics similar to the Sun, has added greater
weight to the idea that magnetic fields in stars may be being
maintained by an interface mechanism described above (Donati
\textit{et al.\ }  2008).

The star that is known as $\tau$ Boo has a similar internal
structure to the Sun, but with a very thin convection zone. The
magnetic field for this star has been monitored over a several year
period and recently the first flip in the large-scale
magnetic field of a star other than the Sun has been seen (Donati \textit{et al.\
} 2008). This is an exciting advance in observations of stellar
magnetic fields and raises questions as to how long it would be
before detect another reversal. In this star, as is the case in the
Sun, there is region of strong shear (a tachocline) where the
convection zone that differentially rotates joins onto the rest of
the interior. This has led the authors to conclude that there is
most likely an interface dynamo mechanism, similar to that for the
Sun discussed above, that is giving rise to the maintenance of a
magnetic field in this star.

As $\tau$ Boo reminds us, there is considerable variation in the
internal structure of stars, and not just in the ratio of the
thicknesses of the layers, which will play an important role in the
internal dynamics between the electrically conducting fluid and the
magnetic field. For some stars, there is a large convectively driven
core as opposed to a large convectively driven envelope just below
the surface. However, there does not have to be only one convective
region within a star and it is believed that there can be
multiple convection zones near the surface of some stars. Such complex
structures inside stars are the result of 
compositional changes as you move radially outward from the centre of the star (see
Silvers \& Proctor 2007 and references there in). Comparing and
contrasting different stars and their magnetic fields in future
years will surely give us greater insight into stellar dynamos.

The precise way in which a magnetic field is maintained in an
astrophysical body, such as the stars mentioned above, is a
fascinating topic and not limited to the consideration of stars. In
fact there has been considerable work to understand the geodynamo --the
dynamo mechanism for the Earth (for a recent review see Kono \&
Roberts 2002). In
recent years, with many missions providing data on planets within
our own solar system, there has been a drive to understand in
greater detail why stars like the Sun, and planets such as the Earth
or Jupiter, show evidence of a dynamo mechanism at work, whilst
Venus has no detectable large-scale magnetic field (see, for
example, Jones 2007 and references therein) and Mars appears to have
have a magnetic field (Stevenson 2001) but there seems to be no
current dynamo mechanism. This has forced
astrophysicists to ask questions about what it is about some planets
that give rise to a working dynamo mechanism.

What we are learning about the magnetic field in the planets in the
solar system will help in the long run with our understanding of
the extra solar planets that are now being detected (see the
Extrasolar Planets Encyclopedia for a list that is frequently
updated\footnote{http://exoplanet.eu/catalog.php}), of which there is
much less detailed observational data. Many of these extra solar
planets are Jovian-like but with considerable variation in the
proximity of the planet to the star it orbits (leading to the name:
hot Jupiters). As such, achieving a comprehensive understanding of
the behaviour of the magnetic field in Jupiter via theory and direct
observations and measurements will help us form a better picture of
these planets that have recently been discovered.

\section{The role of the magnetic field in accretion discs}

Magnetic fields not only play a significant role once stars and
planets are formed, but are also believed to have an important role to play in their
formation due to the interactions of the plasma with the weak
magnetic field that resides in parts of the discs from which these objects are
formed. Matter 
from the disc accretes onto an initially small object
in the centre to form the star or planet.

To accrete onto the central body it is necessary to transport
angular momentum outward. This seems, on the surface, a simple
objective. However, it has given rise to a great deal of debate,
with several mechanisms that give rise to the outward transport of
angular momentum in an accretion disc having been proposed (see
Balbus \& Hawley 1998 for a discussion of the historical
suggestions).

The drive to understand the transport of angular momentum started in
the 1970s with papers such as that of Shakura \& Sunyaev (1973).
Early approaches sought a purely hydrodynamical based reason (see
Balbus \& Hawley, 1998, for a detailed historical overview). Any
mechanism that involved the magnetic field playing a substantial
role was discounted until the early 1990s because the magnetic
field in the accretion discs can be extremely weak and so, as in the
Sun, initially considered to be inconsequential. Hence, it was
believed for many years that what was required was a hydrodynamical
instability yielding turbulence within the disc. Turbulence is known
to give rise to a greatly enhanced transportation and mixing rate of
quantities - for example movement of a drop of dye in a vat of water
is greatly increased over the pure diffusion rate if the water is
turbulent.

However, in the early 1990s it was realised that the role of the
magnetic field had not really been examined despite discussions of the
magnetic field in accretion discs appearing since an article by Lynden-Bell in 1969. It has now become clear that one way to generate turbulence in a disc is by appealing to
an instability that occurs when there is rotation in the presence of
a weak magnetic field (an instability that was first identified in a
different context by Velikov in 1959). Linear stability analysis has show that a
differentially rotating disc, with angular velocity decreasing
outwards, is unstable in the presence of a weak magnetic field
(Balbus \& Hawley 1991; Hawley \& Balbus 1991). This magnetorotational instability
(MRI) gives rise to an enhanced transportation of angular momentum
via the turbulence that is created in the disc.  However, I note here that there are some alternative theories on how accretion
  may be occurring within accretion discs which do involve strong
  magnetic fields (see for example, Ferreira, 1997, for further details).

\begin{figure}[htbp]
\begin{center}
\epsfig{file=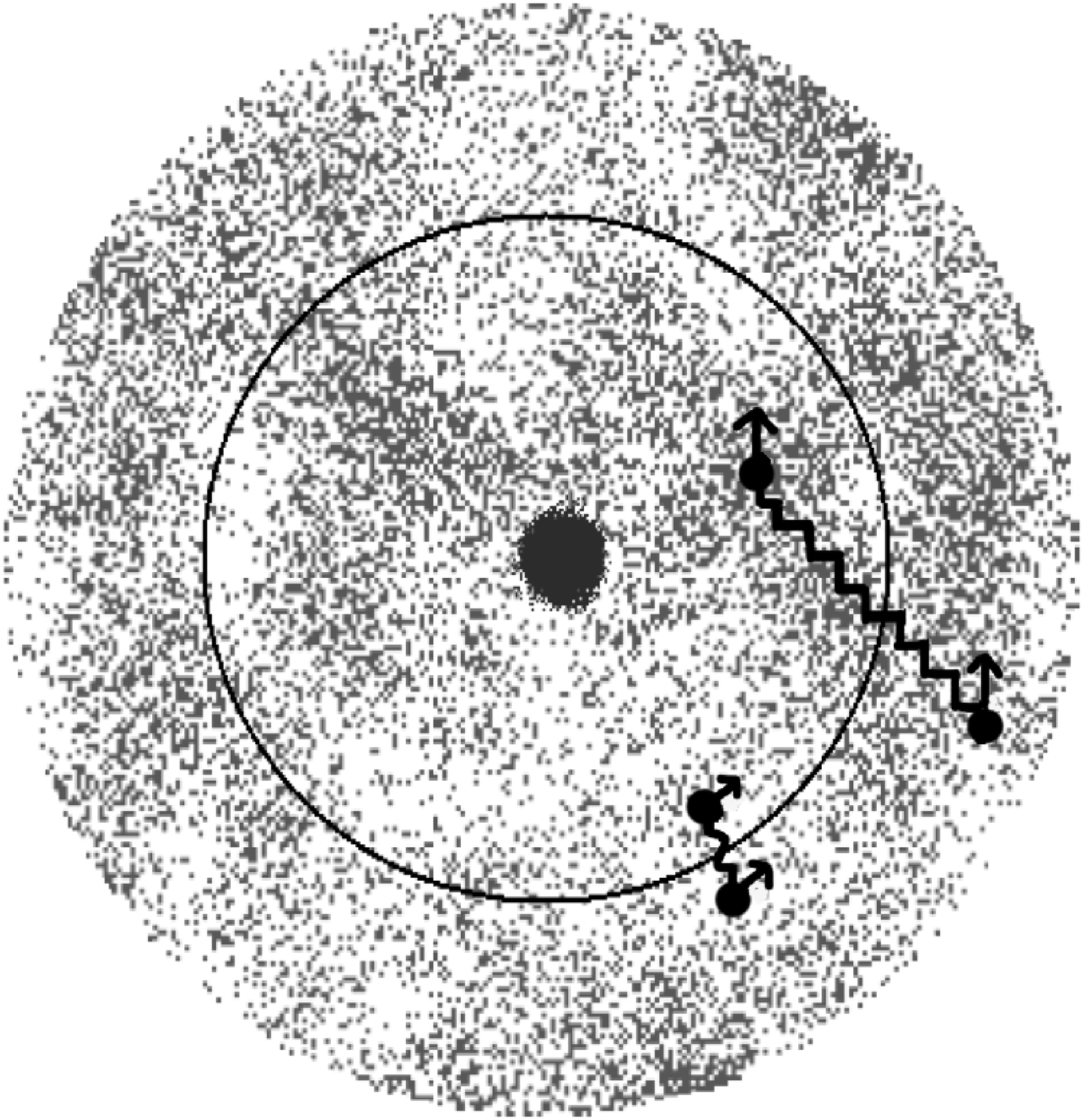,width=7cm,height=5cm}
\end{center}
\caption{Illustration of how the basic MRI instability works in an
  accretion disc.} \label{disc}
\end{figure}

To understand the essence of the MRI instability in a disc, one can
picture the weak magnetic field that resides in the disc as acting
like a spring that tethers two elements in the disc
together\footnote{For simplicity in our toy scenario I  consider the case
 of ideal MHD i.e.\ no viscous etc dissipation.}. Figure
\ref{disc} shows an initial configuration where one of our elements
starts a little closer to the central mass than the other. As such,
the inner element will orbit the central mass at a faster rate,
which would stretch our hypothetical spring that connects them. This
stretching gives rise to a torque that pulls the inner element back
in its orbit and the outer element is pulled forward at the same
time, transferring angular momentum from the inner element to the
outer element. The inner element, which has lost angular momentum,
moves further in, stretching the spring even more, and the process
continues. This descriptive picture of the magnetorotational
instability is only valid for a weak field; if the magnetic field
were strong, the particles would be connected by an extremely stiff
spring or bar, and the run-away instability process cannot occur.
Therefore, for the magnetorotational instability to occur, the
magnetic field must be weak, which is precisely the scenario that is
found in accretion discs.

While there is now a plausible mechanism through which accretion can
occur, it is not clear how efficient it is i.e.\ how fast
angular momentum would be transferred in accretion discs via only this mechanism.
Due to the non-linear coupled equations that govern the system,
determination of this rate requires the use of numerical techniques to
evolve the equations, but with current
resources there is no way that  a fully resolved
simulation of the full disc can be constructed for the viscosity, resistivity and
thermal conductivity values associated with discs.

To
make some progress in the subject, the initial strategy was to
consider the MRI in a small patch of the disc. By reducing
the size of the computational domain, one can obtain higher
resolutions for the same computational cost and so resolve much
smaller scales. Great advances in understanding various issues related
to this instability have been made via local models of a section of
the disc (see, for
example Sano, Inutsuka, Turner \& Stone 2004;  Fromang, Papaloizou, Lesur \& Heinemann 2007; Lesur \& Longaretti 2007; Pessah,
Chan \& Psaltis 2007;
Silvers 2008).  With the computational
resources that exist now, and for the foreseeable future, it is
impossible to work
with the real values for viscosity etc but it is possible to gradually
decrease the values towards more astrophysically relevant values. This
approach may make it possible to determine, for example, the behaviour, and possibly a scaling
law, for how key quantities change as the dissipative parameters are
reduced. However it may also prove helpful to  consider methods to cut
the computational resources even further, e.g.\ by including some
kind of sub-grid scale modelling, but this is a complex issue and
will require considerable thought on the precise way that the
calculations should be carried out. \footnote{For a discussion of some
of the current subgrid scale models see, for example, Buffett (2003).}

It is important to note that, while helpful to some degree, a local modelling approach leaves many
questions unanswered, particularly how the results relate to the full disc
problem. In the full disc there are a lot more questions to resolve
such as the effect of the choice of boundary conditions. As such,
one useful goal for the future, in our way to understanding the rate
of accretion in discs, should be higher resolution computations in full disc
models.

\section{Concluding remarks}

Over the last 100 years, since Hale made the first discovery of a
magnetic field in a non-Earth object, there has been vast progress in
our understanding of the interactions between astrophysical plasmas
and a magnetic field. There is now a detailed picture of how solar
features arise, what ingredients preclude dynamo action in stars and
planets, and how turbulence in a disc gives rise to accretion of
matter onto the central object. This said, there are still a
plethora of questions that remain for us to answer such as:

\begin{itemize}
  \item Why is it the case that there are extended periods in history, such as
  the Maunder minimum
  where there is an absence of the `usual' sunspots that are
  associated with solar cycle? What is the trigger for the sunspot
  cycle to appear again several decades later?
  \item What is it that makes Venus different, such that no
  dynamo mechanism operates?
  \item How does a more complex internal structure within a star affect
  the maintenance and transport of the magnetic field?
  \item Are there further astrophysical areas where the
  magnetic fields have so far been perceived to be weak, and so
  neglected, which now warrant reconsideration in the light of the
  vital role played by weak magnetic fields?
  \item In how many stars is there a flip in the magnetic field, and on
  what time-scales do these flips occur? It is possible to obtain a
  comprehensive understanding of how the time-scale for flips
  relates to the structure of each star?
\end{itemize}

Hopefully in the next 100 years new considerable progress will be made on
these topics and other areas with the help of new space-based
missions and other technology that will improve our current
knowledge from observations, and advanced in computer resources.

 \begin{acknowledgements}
 I would like to thank David Hughes, Michael Proctor, Sam Falle,
 Steve Balbus, Nigel Weiss, John Papaloizou, Paul Bushby, Geoffroy Lesur and Steve
 Tobias for many stimulating conversations over the years.
 \end{acknowledgements}

   \label{lastpage}

\end{document}